\documentstyle[aps,psfig]{revtex}
\begin{document}

\twocolumn[

\begin{center}

{\large\bf Very long transients in globally coupled maps}\\

\vspace{.4cm}
{Susanna C. Manrubia and Alexander S. Mikhailov}\\
{\small \it Fritz-Haber-Institut der Max-Planck-Gesellschaft,
Faradayweg 4-6, 14195 Berlin, Germany}\\
\end{center}

\vspace{.2cm}
{\small Very long transients are found in the partially ordered phase of 
type II of globally
coupled logistic maps. The transients always lead the system in this phase to
a state with a few synchronous clusters. This transient behaviour is not
significantly influenced by the introduction of weak noises. However, such
noises generally favor cluster partitions with
more stable periodic dynamics.
} \\
{PACS number(s): 05.45.-a, 05.45.Xt, 05.40.Ca}
] 

\vspace{0.5cm} 
Globally coupled maps (GCM) formed by ensembles of logistic maps have been
used as a paradigm of complex collective dynamic behaviour for a decade \cite
{KO1,KO2,KDD,MMN,ZL}. Originally, GCM were introduced as a mean-field
approach to coupled map lattices \cite{KCML}, but later the nontrivial
dynamics and the rich collective phenomenology displayed by that system made
it a subject worth of study in itself. One of the main properties of
globally coupled logistic maps is the presence of different phases
characterized by turbulent (non-synchronized) behaviour, clustering, and
global synchronization \cite{KO2}. The formation of a number of subgroups of
synchronized elements out of a symmetrical ensemble has a high relevance for
many applications, such as the organization of the immune or the neural
system, ecological networks, cell differentiation, and structuring of social
hierarchies. Therefore, the GCM phases in which the system displays
clustering have been intensively studied \cite
{KO1,KDD,Xie,glass,Kglass,KanMilnor}.

The former studies about the complex collective behavior displayed by
GCM are usually based on the classical classification of the phase-space
introduced by K. Kaneko \cite{KO2}. As we report here, the partially
ordered phase of type II is equivalent to its neighboring ordered phase,
with the only difference that very long transients preceed the achievement
of the final attractor. This implies a revision of the phase space 
of GCM, and shows that there is a strong non-monotonous 
dependence of the transient length with the system parameters, which has
to be taken into account in any numerical study.

The simplest globally coupled discrete-time system is given by

\begin{equation}
x_{i}(t+1)=(1-\epsilon )f{\bf (}x_{i}(t){\bf )}+{\frac{\epsilon }{N}}%
\sum_{j=1}^{N}f{\bf (}x_{j}(t){\bf )}\;.  
\label{GCM}
\end{equation}
where the individual element evolves according to the logistic map $%
f(x)=1-ax^{2}$, $N$ is the total number of maps and $\epsilon $ specifies
the coupling strength. In general, an attractor of this dynamical system is
formed by a number ${\cal {K}}$ of synchronous clusters each containing $%
N_{k}$ elements, $k=1,\dots {\cal {K}}$ and can be characterized by means of
the partition $({\cal {K}};\,N_{1}\ge N_{2}\ge \dots \ge N_{{\cal {K}}})$.
For convenience,  this classification also includes one-element ''clusters''
($N_{k}=1$) that actually correspond to individual non-entrained elements.
Thus, the partition ($N;1,1,...,1$) corresponds to the asynchronous state of
the entire ensemble, while the partition ($1;N$) represents its fully
synchronous state, where all $N$ elements belong to a single cluster. In
addition to these two states, the system would generally also have other
partitions where a certain number of synchronous groups of elements with $N>%
{\cal {K}}>1$ are present. The choise of an attractor with a particular
partition is determined by the initial conditions. The attractor
corresponding to each initial condition is characterized by a certain number
${\cal {K}}_{m}$ of clusters after a transient has elapsed.

The phase space of the GCM (\ref{GCM}) has been described by K. Kaneko 
\cite{KO2} using the average cluster number ${\overline{{\cal {K}}}}
=\sum_{m=1}^{M}{\cal {K}}_{m}/M,$ where the index $m=1,\dots ,M$ enumerates
the set of employed initial conditions. Four different phases have been
identified \cite{KO2}:

\begin{enumerate}
\item  {\it Coherent phase.} The elements follow the same trajectory ($%
x_{i}(t)=x_{j}(t)$, $\forall i,\;j;\;t$), forming a single synchronous
cluster ($\overline{{\cal {K}}}=1$).

\item  {\it Ordered phase.} Almost all basin volume is occupied by a
few-cluster attractor ($\overline{{\cal {K}}}$ is small and does not grow
with $N$).

\item  {\it Partially ordered phase.} Coexistence of many-cluster and
few-cluster attractors ($\overline{{\cal {K}}}$ is large and grows with 
$N$).

\item  {\it Turbulent phase.} No synchronization among the elements 
($\overline{{\cal {K}}}=N$).
\end{enumerate}

The coexistence of many-cluster and few-cluster attractors has been observed
by K. Kaneko in two different parameter intervals. One of them separated the
ordered and the turbulent phases. Here the system is in the partially
ordered phase of type I, also called the {\it intermittent phase}. The other
interval lies between the regions occupied by the ordered and coherent
phases. In this interval the system is in the partially ordered phase of
type II (called the ''glassy'' phase in the initial publication \cite{KO2}).
The typical parameter intervals are $1.56<a<1.80$ for $\epsilon =0.3$
(partially ordered phase of type II) \cite{KO2,KanInf}, and $1.58<a<1.69$
for $\epsilon =0.1$ (intermittent phase) \cite{KO2,glass,Kglass,KanInf}.

To compute the asymptotic properties of a dynamical system, one has to
ensure that the system has had enough time to approach its final state, i.e.
that the dynamical attractor for the given initial conditions has been
reached. Slow relaxation is indeed known for some dynamical systems (see,
e.g., \cite{Daido}). The properties of the transients of GCM have not yet
been sufficiently investigated. The aim of the present Letter is to
systematically study the transient behaviour of GCM, described by equation (%
\ref{GCM}). Our principal result is that inside the whole parameter region,
corresponding to partially ordered (''glassy'') phases of type II, only
few-cluster attractors are observed after very long transients. This result
holds also when weak noises are added. It contradicts to what has previously
been reported by K. Kaneko \cite{comment1}.

We performed long runs of up to $T=10^{7}$ iterations and recorded the time
at which the final minimum value of ${\cal K}$ was reached in the parameter
region corresponding to the partially ordered phase of type II. To this end,
the partition $({\cal {K}};\,N_{1}\ge N_{2}\ge \dots \ge N_{{\cal K}})$ was
determined every $\Delta t=25-50$ time steps. Double precision real numbers
were used in these computations, ensuring the absolute precision of $10^{-16}
$. Two elements were taken to belong to the same cluster only if they had
exactly the same state within the double computer precision, i.e. if $%
|x_{i}(t)-x_{j}(t)|<10^{-16}$.

\begin{figure}[tbp]
\label{trans} \centerline{\psfig{file=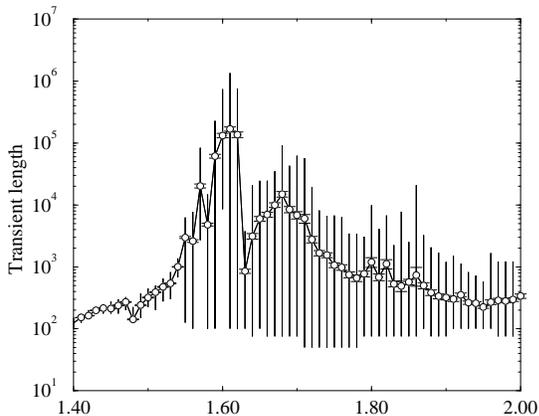,width=8.cm,angle=270}}
\caption{Average time required to approach an attractor as function of the
control parameter $a$ for the coupling strength $\epsilon=0.3$ in a system
of size $N=100$. Averages over 100 different initial conditions are
performed. Vertical bars indicate minimum and maximum values of the 
transient time.}
\end{figure}

The computed average transient length for $\epsilon =0.3$ as a function of $a
$ is shown in Figure 1. We see that the transients may extend up to tens of
thousands and even millions of time steps. They become especially long near $%
a=1.6$. Previous numerical studies \cite{KO1,KO2,KanInf} were limited to
much shorter evolution times (up to $10^{4}$ iterations) and therefore some
of the behaviour observed in these studies essentially corresponded to
transients. This becomes clear if we compare our Fig. 1 with Fig. 9 in Ref.
\cite{KO2}: A strong increase in the mean number of clusters $\overline{%
{\cal {K}}}$ was reported exactly where the transient length greatly
increases (exceeding $3000$ time steps). Our investigation reveals that,
after long transients, only attractors with ${\cal {K}}\le 2$ are typically
found for $a<1.65$, and only attractors with ${\cal {K}}\le 6$ are observed
for $1.65<a\le 2$. Similar results are also obtained in our calculations for
$\epsilon =0.25,\;0.35,$ and $0.4$.

\begin{figure}[tbp]
\label{decay} \centerline{\psfig{file=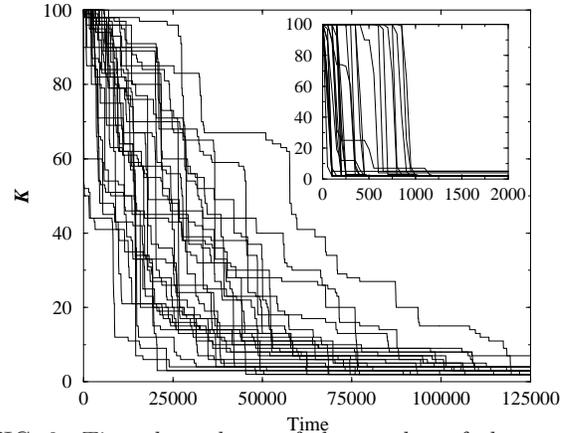,width=8.cm,angle=270}}
\caption{Time dependence of the number of clusters in a system of size $N=100
$ for the coupling intensity $\epsilon=0.3$ at $a=1.6$ (main plot) and $a=1.8
$ (inset), and 30 different initial conditions in each case. }
\end{figure}

The time dependence of the number of clusters ${\cal K}(t)$ at $\epsilon =0.3
$ for 30 different initial conditions is shown in Figure 2 for $a=1.6$ (main
plot) and $a=1.8$ (inset). For $a=1.6$, the number of clusters is indeed 
large during the initial evolution, and comparable with the total size of the
system ($N=100$). Later on the number of clusters is slowly decreasing and
eventually only attractors with ${\cal {K}}=2$ (but different partitions $%
N_{1},\;N_{2}$) are found at this value of the parameter $a$. The system
evolution at $a=1.8$ is essentially similar, though it is characterized by a
much faster convergence to the final states (note the difference by almost
two orders of magnitude in the time scales in these two plots). Another
difference is that for $a=1.8$ the final states with various cluster numbers
${\cal {K}}=2,\;3$ and $4$ are observed. By averaging over a large number of
initial conditions, the time dependence for the relaxation of the mean
cluster number ${\overline{{\cal K}}}(t)$ to its asymptotic value ${\cal {K}}$ 
has been obtained. Figure 3 shows in the logarithmic scale the
time dependence of the quantity $\delta {\cal {K}}=$ ${\overline{%
{\cal {K}}}}(t)-{\cal K}$ for $\epsilon =0.3$ and $a=1.6$. We
clearly see that the relaxation is exponential, $\delta {\cal{K}}
\propto \exp \{ -\beta t \}$.

We have further analysed how the mean transient time $\tau =\beta ^{-1}$
depended on the system size $N$. The explored interval of system sizes was $%
2^{4}\le N\le 2^{12}$; we have used several values of $a$ and fixed the
coupling strength at $\epsilon =0.3$. We did not find any strong variation
of $\tau $ with $N$, i.e. the order of magnitude of $\tau $ did not depend
on the system size. The transient length depicted in Fig. 1 is
characteristic for almost three decades of variation in the system size $N$.

\begin{figure}[tbp]
\label{expdecay} \centerline{\psfig{file=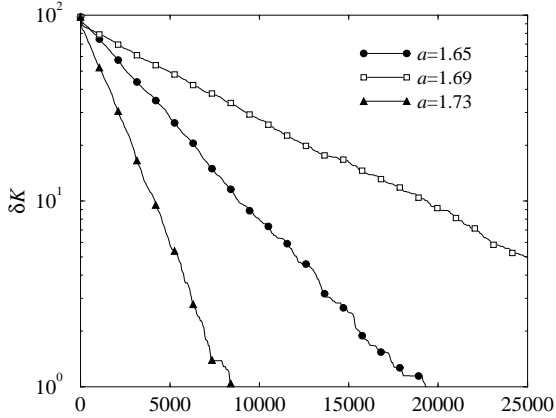,width=8.cm,angle=270}}
\caption{Exponential relaxation to the asymptotic state for the coupling
intensity $\epsilon=0.3$ and three different values of the control parameter
$a$. The slope of the curves (from top to bottom) is $\beta=1.1 \times 10^{-4}
$, $2.4 \times 10^{-4}$, and $5.4 \times 10^{-4}$, and errors are of order 
$10^{-7}$. The system size is $N=100$. Averaging over $M=10^3$ initial 
conditions is performed. }
\end{figure}

The presence of very long transients indicates that the system may be
sensitive to the application of noise. Indeed, for the so-called Milnor
attractors even a tiny perturbation would suffice to destabilize the
asymptotic state \cite{comment}. The existence of Milnor attractors has been
discussed both for the partially ordered phase of type II \cite{KanInf} and
for the intermittent phase \cite{KanMilnor}. To analyze the effect of weak
random perturbations, we have modified equations (\ref{GCM}) by adding a
noise term $\eta \;r_{i}(t)$. We have chosen a small noise intensity $\eta
=10^{-10}$; independent random numbers $r_{i}(t)\in (-1,1)$ are drawn anew
from a uniform distribution for each element and at each time step. Noise
prevents the spurious synchronization of elements in the system: If the
states of two maps $i$ and $j$ are equal (with computer precision) at time $t
$, they will follow identical trajectories for all $t^{\prime }>t$ in a pure
deterministic system. When noise is added, spurious attractors are not
attained and only robust attractors should be detected.

In the presence of noise the states of elements in a cluster cannot be
identical. To define a cluster, we have to choose a certain finite precision
$\delta $ and say that elements $i$ and $j$ belong to the same cluster at
time $t$ if $|x_{i}(t)-x_{j}(t)|<\delta $ (cf. the respective definition for
the case of randomly coupled maps \cite{MMN}). We have found that the
application of weak noise does not qualitatively influence the
above-described evolution. Figure 4 shows in logarithmic scale the mean
number ${\overline{{\cal {K}}}}(t)$ of clusters as function of time when
weak noise is present (all other parameters are the same as in Fig. 3).
The system still evolves towards final distributions characterized by a few
large clusters. Typically, a slower convergence to the limit value 
${\cal {K}}$$=2$ was observed when the noise was acting. Only in a very
narrow domain $1.60 \le a \le 1.62$ did noise seem to prevent
convergence to a few-cluster attractor. Note that this area coincides
with the maximum transient length in the deterministic case, and is near
the boundary where the single synchronous cluster becomes
unstable (it is known that the unstabilization of the coherent phase
proceeds through a power-law divergence of the transient lenght
\cite{MMN}).

\begin{figure}[tbp]
\label{Tnoise} \centerline{\psfig{file=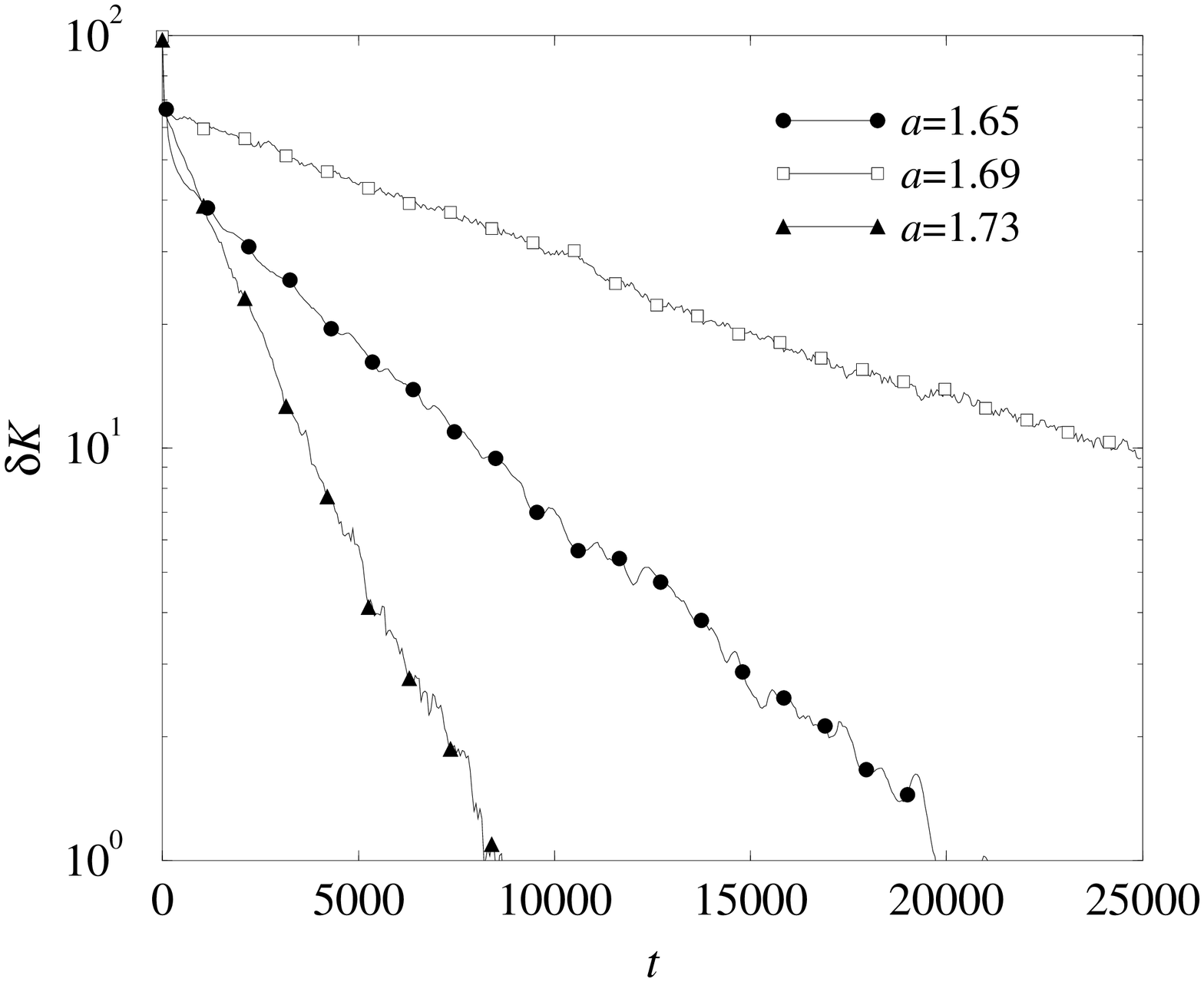,width=8.cm,angle=270}}
\caption{Exponential relaxation in GCM added with
noise of amplitude $\eta = 10^{-10}$ (other parameters as in Fig. 3). The 
precision is $\delta=\eta^{1/2}$. The
slope of the curves (from top to bottom) is $\beta= 8 \times 10^{-5}, \; 2
\times 10^{-4}$, and $5 \times 10^{-4}$. }
\end{figure}

The dynamics corresponding to a particular cluster partition in our
simulations was either periodic or (intrinsically) chaotic. To detect
intrinsically chaotic dynamics, local Lyapunov exponents were examined.
After a fixed transient of length $T=10^{7}$ we numerically calculated the
local Lyapunov exponent

\begin{equation}\label{LY}
\lambda _{m}(\epsilon ,a)={\frac{1}{N\tilde{t}}}\sum_{t=T}^{T+{\tilde{t}}%
}\sum_{j=1}^{N}\log |f^{\prime }(x_{j}(t))|
\end{equation}
corresponding to the trajectories $x_{j}(t)$ of elements $j$ for the given
initial condition $m$ and parameters $\epsilon $ and $a$. The averaging time
was always ${\tilde{t}}=10^{4}$. Positive exponents correspond to chaotic
dynamics. The same procedure was used both in the presence and in absence of
the noise.

When noise is acting, it may, in principle, induce transitions from one
cluster partition to another. Our numerical investigations show, however,
that such transitions actually take place in the presence of very weak
noises only if the dynamics corresponding to a particular cluster partition
is intrinsically chaotic. This observation leads us to a conjecture that
Milnor attractors in GCM are, perhaps, only generated by cluster partitions
with intrinsically chaotic dynamics. Cluster partitions with periodic
attractors are stable against a finite amount of perturbation, while the
system leaves with certainty a partition with a chaotic attractor in a
finite time when noise is present (see also \cite{KanMilnor}).

\begin{figure}[tbp]
\label{fnoise} \centerline{\psfig{file=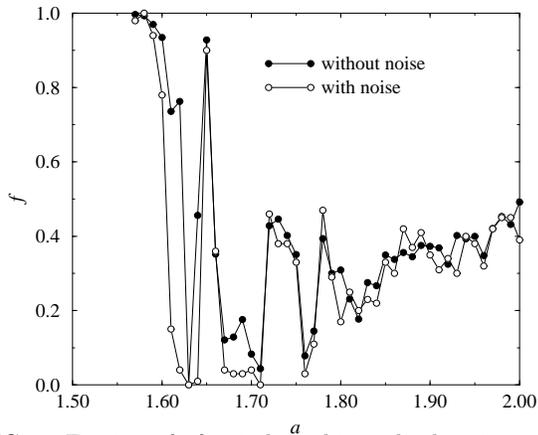,width=8.cm,angle=270}}
\caption{Fraction $f$ of initial conditions leading to a chaotic attractor
for GCM with and without noise. The parameters are $\epsilon =0.3$ and $N=100
$ and we have averaged over $M=10^{3}$ initial conditions.}
\end{figure}

Interestingly, the addition of noise favors the attainement of periodic
attractors. Figure 5 shows the fraction $f$ of initial conditions leading to
a cluster partition with (intrinsically) chaotic dynamics with and without
noise for the coupling intensity $\epsilon =0.3$, as identified by means
of (\ref{LY}). We see that this fraction
is strongly reduced in the presence of noise around the parameter value $%
a=1.61$. To explain this, suppose that the system has approached a cluster
partition with chaotic dynamics. Given that any such partition is
destabilized even by weak noise, we expect that elements would spend only
some time near this attractor, but then one of them would change its cluster
affiliation and a new cluster partition would thus be produced. As long as
this new partition is also chaotic, the system again easily escapes and the
same procedure repeats until a much more stable partition with periodic
dynamics is found. This simple argument predicts that, under the action of
noise, the system would wander between chaotic Milnor attractors until it
eventually finds a robust periodic attractor. If this is indeed so, chaotic
attractors would always represent mere transients for sufficiently weak
noises. Nonetheless, to test such hypothesis much longer iterations are
apparently needed.

Thus, our numerical analysis of globally coupled logistic maps has shown
that the collective dynamics of this system in the partially ordered phase
of type II is characterized by the presence of very long transients. The
asymptotic states of the system in this parameter region are, however, the
same as in the ordered phase and include only a small number of synchronous
clusters. This conclusion holds even when a small noise, eliminating spurious
attractors, is introduced. We have also performed the analysis of dynamical
transients in the intermittent phase, i.e. at the interface between the
ordered and the turbulent phases, and have found (to be separately
published) that in this region the coexistence of few- and many-cluster
attractors is indeed observed. These results are important for the
general classification of dynamical behaviour in GCM.

The authors acknowledge the financial support from the
Alexander-von-Humboldt Foundation (Germany).


\begin{thebibliography}{99}
\bibitem{comment1}  The possibility of very long transients in this region
has been mentioned by K. Kaneko in one of his publications \cite{Kglass},
but not further explored.
\end{thebibliography}
\end{document}